%% file: main.tex
\documentclass[10pt,conference]{IEEEtran}
\IEEEoverridecommandlockouts
\input{preamble}

\usepackage{cite}
\usepackage{amsmath,amssymb,amsfonts}
\usepackage{algorithmic}
\usepackage{graphicx}
\usepackage{textcomp}
\usepackage{fontawesome}
\usepackage{makecell}
\usepackage{balance}
\usepackage{csquotes}
\usepackage{tikz}

\def\BibTeX{{\rm B\kern-.05em{\sc i\kern-.025em b}\kern-.08em
		T\kern-.1667em\lower.7ex\hbox{E}\kern-.125emX}}
\begin{document}
	
	\title{Detecting Connectivity Issues in Android Apps}

	\author{\IEEEauthorblockN{Alejandro Mazuera-Rozo\textsuperscript{1,2}, Camilo Escobar-Vel{\'a}squez\textsuperscript{2}, Juan Espitia-Acero\textsuperscript{2}\\Mario Linares-V{\'a}squez\textsuperscript{2}, Gabriele Bavota\textsuperscript{1}}
		\IEEEauthorblockA{\textsuperscript{1}\textit{SEART} @ Software Institute, Universit\`a della Svizzera italiana, Lugano, Switzerland \\
			\textsuperscript{2}\textit{The Software Design Lab} @ Universidad de los Andes, Bogot\'a, Colombia 
		}
	}	
	
	\maketitle
	
	\thispagestyle{empty}
	\pagestyle{empty}
	
	\begin{abstract}
Android is the most popular mobile operating system in the world, running on more than 70\% of mobile devices. This implies a gigantic and very competitive market for Android apps. Being successful in such a market is far from trivial and requires, besides the tackling of a problem or need felt by a vast audience, the development of high-quality apps. As recently showed in the literature, connectivity issues (\eg mishandling of zero/unreliable Internet connection) can result in bugs and/or crashes, negatively affecting the app's user experience. While these issues have been studied in the literature, there are no techniques able to automatically detect and report them to developers. We present \approach, a tool able to detect statically 16 types of connectivity issues affecting Android apps. We assessed the ability of \approach to precisely identify these issues in a set of \numapps open source apps, observing an average precision of \CONANPrecision\%. Then, we studied the relevance of these issues for developers by (i) conducting  interviews with six practitioners working with commercial Android apps, and (ii) submitting \numissues issue reports for \openSource open source apps. Our results show that several of the identified connectivity issues are considered as relevant by practitioners in specific contexts, in which connectivity is considered a first-class feature.
	\end{abstract}
	

\input{secs/introduction}
\input{secs/related_work}
\input{secs/approach}

\input{secs/study_design}
\input{secs/results}

\input{secs/threats}
\input{secs/conclusion}

\section{Acknowledgements}
	Mazuera-Rozo and Bavota gratefully acknowledge the financial support of the Swiss National Science Foundation for the CCQR project (SNF Project No. 175513). Escobar-Vel\'asquez and Linares-V\'asquez were  funded by a Google Latin America Research Award 2018-2021. Escobar-Velásquez was supported by a ESKAS scholarship, No. 2020.0820.

\clearpage
\balance
\bibliographystyle{IEEEtran}
\bibliography{bib/main}

\end{document}

%% file: preamble.tex
\usepackage[hidelinks]{hyperref}
\usepackage[plain]{fancyref}
\usepackage{ifdraft}

\usepackage[inline]{enumitem}
\usepackage{xcolor}
\usepackage{xspace}
\usepackage[final]{listings}
\usepackage{acronym}
\usepackage{url}
\usepackage{amsmath}
\usepackage{booktabs} 
\usepackage{subfig}
\usepackage{balance}
\usepackage{dirtree}
\usepackage{multirow}
\usepackage{wasysym}

\usepackage[ruled]{algorithm2e} 

\usepackage{etoolbox}
\makeatletter
\patchcmd{\@makecaption}
  {\scshape}
  {}
  {}
  {}
\patchcmd{\@makecaption}
  {\\}
  {.\ }
  {}
  {}
\makeatother

\definecolor{OliveGreen}{rgb}{0,0.6,0.3}

\def\fref{\Fref} 
\renewcommand{\lstlistingname}{Snippet}
\newcommand*{\fancyreflstlabelprefix}{lst}
\newcommand*{\Freflstname}{\lstlistingname}
\newcommand*{\freflstname}{\MakeLowercase{\lstlistingname}}
\Frefformat{vario}{\fancyreflstlabelprefix}%
  {\Freflstname\fancyrefdefaultspacing#1#3}
\frefformat{vario}{\fancyreflstlabelprefix}%
  {\freflstname\fancyrefdefaultspacing#1#3}
\Frefformat{plain}{\fancyreflstlabelprefix}%
  {\Freflstname\fancyrefdefaultspacing#1}
\frefformat{plain}{\fancyreflstlabelprefix}%
  {\freflstname\fancyrefdefaultspacing#1}

\renewcommand{\tablename}{Table}  
  
\newcommand*{\fancyreflnlabelprefix}{ln}
\newcommand*{\Freflnname}{Line}
\newcommand*{\freflnname}{\MakeLowercase{\Freflnname}}
\Frefformat{vario}{\fancyreflnlabelprefix}%
  {\Freflnname\fancyrefdefaultspacing#1#3}
\frefformat{vario}{\fancyreflnlabelprefix}%
  {\freflnname\fancyrefdefaultspacing#1#3}
\Frefformat{plain}{\fancyreflnlabelprefix}%
  {\Freflnname\fancyrefdefaultspacing#1}
\frefformat{plain}{\fancyreflnlabelprefix}%
  {\freflnname\fancyrefdefaultspacing#1}

\lstdefinelanguage{JavaScript}{
keywords={typeof, new, true, false, catch, function, return, null, catch, switch, var, if, in, for, while, do, else, case, break, throw, this, instanceof},
keywordstyle=\color{purple}\bfseries,
ndkeywords={},
ndkeywordstyle=\color{blue}\bfseries,
identifierstyle=\color{black},
sensitive=false,
comment=[l]{//},
morecomment=[s]{/*}{*/},
commentstyle=\color{OliveGreen}\ttfamily,
stringstyle=\color{OliveGreen}\ttfamily,
morestring=[b]',
morestring=[b]"
}
\usepackage{color}
\definecolor{gray97}{gray}{.97}
\definecolor{gray90}{gray}{.90}
\definecolor{gray75}{gray}{.75}
\definecolor{gray45}{gray}{.45}
\definecolor{codegreen}{rgb}{0,0.6,0}
\definecolor{codegray}{rgb}{0.5,0.5,0.5}
\definecolor{codepurple}{rgb}{0.58,0,0.82}
\lstdefinestyle{floating}{%
  frame=none,
  float=htb,
  captionpos=b
}

\lstdefinestyle{ctxtraits}
 {language=JavaScript,
  frame=lines,
  showstringspaces=false,
  keywordstyle=\tt\bf,
  tabsize=3,
  style=floating,
  morekeywords={Trait, cop, Context, activate, deactivate, adapt, addObjectPolicy, manager}
}

\lstnewenvironment{ctxtraits}[1][]
 {\lstset{style=ctxtraits,#1}}{}

\newcommand{\eg}{\emph{e.g.,}\xspace}
\newcommand{\ie}{\emph{i.e.,}\xspace}
\newcommand{\etc}{\emph{etc.}\xspace}
\newcommand{\etal}{\emph{et al.}\xspace}

\newcommand{\secref}[1]{Section~\ref{#1}\xspace}
\newcommand{\figref}[1]{Fig.~\ref{#1}\xspace}

\newcommand{\tabref}[1]{Table~\ref{#1}\xspace}
\newcommand*\circled[1]{\tikz[baseline=(char.base)]{
		\node[shape=circle,draw,inner sep=0.3pt] (char) {#1};}}

\newcommand{\eci}{\emph{connectivity issues}\xspace}
\newcommand{\nt}{\emph{NT}\xspace}
\newcommand{\approach}{\texttt{\textbf{CONAN}}\xspace}
\newcommand{\numapps}{44\xspace}

\newcommand{\numissues}{84\xspace}
\newcommand{\CONANPrecision}{80}
\newcommand{\openSource}{27\xspace}

\newcommand{\candidates}{898\xspace}
\newcommand{\manual}{389\xspace}

\newcommand{\code}[1]{\texttt{#1}}

\definecolor{author}{rgb}{.5, .5, .5}
\definecolor{comment}{rgb}{.1, .0, .9}
\definecolor{note}{rgb}{.9, .4, .0}
\definecolor{idea}{rgb}{.1, .7, .0}
\definecolor{missing}{rgb}{.9, .1, .0}
\definecolor{deleteme}{rgb}{.9, .1, .0}

\input{acronyms}

%% file: acronyms.tex
\acrodef{APK}{Android Application Package}

%% file: secs/introduction.tex

\section{Introduction}
The always increasing popularity of mobile devices such as smartphones has triggered a strong growth of the mobile apps market in recent years. These apps support everyday activities such as online shopping, social networking, banking, \etc It comes without surprise that the research community put the attention on studying mobile apps from different perspectives \cite{Nagappan:saner2016,Martin:tse2017}. For example, several studies highlighted the role played by the apps' code quality (\eg usage of robust APIs) on their success \cite{Bavota:tse2015, Linares:FSE13, Tian:icsme2015,Noei:emse2017}. Among the several quality attributes that are relevant for the user experience, those related to Internet connectivity are becoming more and more important\footnote{Note that in this work we use the term ``connectivity'' to refer to the Internet connectivity. Other types of connectivity (\eg bluetooth, NFC) are out of the scope of this paper.}. Indeed, as it has been shown in the literature \cite{bakar:permissions,Frank2012,Mostafa2017,olmstead2015apps}, network permissions are the most requested by Android apps, showing the central role Internet access plays for their functioning.

A recent work by Escobar-Vel\'asquez \etal~\cite{escobar:emse2021} studied connectivity issues affecting Android apps, namely suboptimal implementation choices that affect the user experience in \emph{zero/unreliable connectivity scenarios}. For example, an application triggering a network operation without firstly checking the availability of Internet connection can crash in case of no connectivity \cite{escobar:emse2021}. 

The reported findings confirmed the prevalence of these issues in 50 open source apps analyzed by the authors, who reported a total of 304 connectivity issues present in them.

Building on top of the work by Escobar-Vel\'asquez \etal~\cite{escobar:emse2021}, we present in this paper \approach, the first approach able to automatically detect \texttt{\textbf{CON}}nectivity issues in \texttt{\textbf{AN}}droid apps. We designed \approach in multiple steps. First, we studied the related literature (\eg \cite{escobar:emse2021, googleGuidelines, googlePermissions, Jha2017, archibald202, EscobarTSE2020, googleThreads, workManager, hellman2014android, connForBillions}) and online resources such as the official Android development guidelines \cite{androidDevGuide} to define a catalogue of connectivity issues to target. From this analysis, we distilled 16 connectivity issues  ($CI$) that \approach is currently able to detect by using static analysis.

Then, we defined for each of these issues $CI_i$ a detection strategy, namely a set of rules that can be checked via static analysis to identify $CI_i$'s instances. The definition of these rules was not straightforward. Mobile apps mostly rely on external libraries (\eg OkHttp \cite{okhttp}) to handle their network (HTTP) requests. This means that the implementation of a detection strategy must be instantiated to support specific libraries. Indeed, the implementation patterns change from library to library, with different APIs involved. For this reason, we ran a survey with 98 Android developers asking which HTTP libraries they mostly use when building apps. Based on the results, we implemented full support for Retrofit~\cite{retrofit}, OkHttp \cite{okhttp}, Volley \cite{volley}, and HttpURLConnection \cite{HttpURLConnection}, which were the main libraries reported by developers.

Once we built our tool, we ran it on a set of \numapps open source apps, identifying \candidates candidate connectivity issues in them. Two of the authors manually inspected a statistically significant sample (\manual instances) of these candidates, to classify them as true or false positives. This evaluation provided us with information about the precision of the detection rules we implemented that, on average, resulted to be \CONANPrecision\%.

Finally, we assessed the relevance of the issues identified by \approach.  First, we interviewed six practitioners and collected their opinions about the relevance of the 16 issue types we detect. The focus here was on the type of issue in general rather than on specific  instances identified by \approach. Second, we opened \numissues issue reports for \openSource open source apps, reporting some of the true positive instances we identified. The study participants indicated that the relevance of the identified issues strongly depends on the specific type of app in which they are identified.  The results of the two studies highlight that, despite all 16 issues we support represent sub-optimal implementation choices for handling connectivity in Android apps, some of them are not perceived as relevant by practitioners. 

\approach implementation and all data used in our study are publicly available \cite{replication}.

%% file: secs/related_work.tex

\section{Related Work}


Techniques and tools have been proposed in the literature to improve different quality attributes of mobile apps \cite{LI201767}, such as: performance-related concerns \cite{ParsonsM08,TrubianiBHAK18, GrechanikFuXie2012, liu2014characterizing}, (ii) security issues \cite{Sadeghi:ICSE15,Bagheri:TSE15,Ahmad:MSR16,Sadeghi:TSE16,bello2019}, (iii) poor responsiveness \cite{Yang2013, Xiong2018,Zhao2019}, and (iv) energy consumption \cite{Linares:MSR14,Linares:TOSEM18,Li2013ISSTA,Behrouz2015,Hao2013}. Most of these approaches rely on static and dynamic analysis, while more recent proposals try to learn semantic and syntactic features of problematic code components by using deep learning models. 

Despite the fact that Internet connectivity is becoming increasingly important for mobile apps, quality issues related to it have not been addressed by any of the previous approaches  \cite{LI201767}. However, related to the ``connectivity topic'', previous works investigated issues related to network profiling \cite{Nayam2016,Mostafa2017,Conti2018,Rapoport2017}, network privacy concerns \cite{Cheng2017,continella17,Huang2019,Gadient2020}, and network exploitability \cite{Shabtai2011, Crussell2014,Wei2012a}. Also, while no tool has been specifically designed for detecting \eci, some of the techniques proposed in the literature to identify apps' crashes, can detect crashes caused by a mishandling of the lack of connectivity. For instance, Moran \etal introduced CrashScope \cite{Moran:ICSE17,Moran:ICST16}, a tool to test mobile apps by employing (i) static and dynamic analysis, and (ii) automatically generated gestures and contextual inputs. Similarly, Adamsen \etal \cite{Adamsen:ISSTA15} and Azim \etal \cite{Azim14} proposed tools for the automated exploration of apps, with the goal of identifying crash events. The aforementioned approaches systematically explore the app to test and generate events like intermittent network connectivity. However, these tools are limited in the type of connectivity issue they investigate (\eg intermittent connectivity).

Liang \etal presented Caiipa \cite{Liang2014}, a mobile app testing framework employing contextual fuzzing methods to test Android apps over an expanded context space. Caiipa is able to simulate a variety of contexts observed in the wild including diverse network-related factors, for example, the throughput and loss rates observed for Wi-Fi and 3G networks. Similarly, JazzDroid \cite{Xiong2018} is an automated fuzzer which injects into applications various environmental interferences, such as network delays, network traffic interference and HTTP requests manipulation. While these tools address a variety of bugs, they are not specialized to \eci. 

More broadly, there are methods for diagnosing poor responsiveness in Android apps given some connectivity shortcomings. Relevant tools under this category are Monkey \cite{MONKEY}, AppSPIN \cite{Zhao2019}, TRIANGLE \cite{Panizo2019} and the approach proposed by Yang \etal \cite{Yang2013}. These tools could eventually test Android apps looking for \eci, by assuming that random events could lead to \eci.

\approach differs from current approaches since our line of action is aligned to static analysis mechanisms. \approach does not need the app to be run nor the simulation of flaky Internet connectivity. \approach allows practitioners to spot potential \eci in the early development stages, since all it needs is the source code of the app to analyze.

%% file: secs/approach.tex

\section{Approach overview} \label{sec:approach}

\approach is a static analysis tool built on top of Android Lint~\cite{AndroidLint}, and aimed at identifying \eci in Android apps. \approach is able to detect these issues for apps that use the libraries described in \secref{sub:suppLibraries}. The list of \eci that \approach supports has been defined based on: (i) official resources for Android developers (\ie Google developer guidelines); (ii) library-specific documentation; and (iii) research articles on this topic. 

\subsection{Android Lint Overview}
\label{sub:androidLint}

Besides ensuring that an app meets its functional requirements, it is also important to guarantee its quality (\eg the absence of bugs). Therefore, the Android SDK provides a tool named Lint that is able to identify code quality issues via static analysis. Android Lint is capable of identifying over 400 issue types \cite{lintIssueIndex} 
corresponding to a variety of quality attributes such as accessibility, correctness, performance, and security. Android Lint follows the open-closed principle, allowing to add custom rules that can detect quality issues of interest.

Android Lint supports the official programming languages used in Android, \ie Kotlin and Java. While Java has been historically the official programming language for creating Android apps, Kotlin is currently the recommended language for native Android apps. By exploiting an universal abstract syntax tree (UAST), Android Lint can analyze source code written both in Java and Kotlin, since both of them are represented using the UAST while running the static analysis. Thus, it is sufficient to write the detection rule for a specific quality issue and Lint provides support for both languages. Additionally, the Android Lint is also IDE-independent, meaning that it can be run through the command line either as a standalone tool or by employing the Gradle wrapper \cite{GradleWrapper} 
present in Gradle-based Android projects. Given these characteristics, we integrated the detection of \eci in Android Lint. 

\subsection{Supported libraries}
\label{sub:suppLibraries}

As explained before, the implementation of the detection strategies strongly depends on the network libraries used by developers. To decide the libraries for which \approach should provide support, we ran a survey investigating which libraries Android developers use to handle network requests. The survey has been designed to last at most 5 minutes to maximize the completion rate and it was structured as reported in \tabref{tab:survey}. We collected background information about participants ($Q_1$-$Q_4$). If a participant answered ``zero'' to $Q_4$ (\ie no  experience with native Android apps) the survey ended and the participant was excluded from the study (4 cases). $Q_5$ aimed to collect information about the libraries developers use to implement network requests in Android apps. We provided a predefined list of existing libraries, with the possibility of specifying additional libraries. The predefined list included, among others, well-known libraries such as: \emph{Retrofit}, \emph{OkHttp}, and \emph{Volley}. 

\begin{table}[ht]
	\scriptsize
	\centering
	\captionsetup{justification=centering}
	\caption{Survey on network libraries used in Android apps}
	\label{tab:survey}
	\resizebox{\linewidth}{!}{
		
		\begin{tabular}{p{6.5cm}}
			
			\toprule
			
			\textbf{BACKGROUND QUESTIONS} \\\midrule
			$Q_1$: In which country do you live?\\
			$Q_2$: What is your current job position?\\
			$Q_3$: How many years of programming experience do you have?\\
			$Q_4$: How many years of programming experience do you have concerning native Android apps?\\\midrule
			\textbf{ADOPTED NETWORK LIBRARIES} \\\midrule
			$Q_5$: Which libraries do you use to implement network requests in Android apps?\\\bottomrule	
			
		\end{tabular}
		
	}
\end{table}

To avoid a leading answer bias, each participant was presented with a shuffled version of the predefined list. 

In addition, an option stating ``\emph{I have never used an API for network request}'' was included.

Besides sharing the survey with developers from companies we know, it was also shared in social media and within the Google Android developers group \cite{AndroidDevelopers}. 
We collected answers for two months, with a total of 98 participants that completed our survey from 41 countries (\eg Germany, USA, Ukraine, \etc). On average, participants had $\sim$10 years of programming experience and $\sim$5 years of Android development experience. Regarding their job position, 7\% of participants are B.Sc. students, 16\% M.Sc. students, 2\% Ph.D students, 1\% Postdoc and 74\% professional software engineers having different specializations (\eg CTO, CoFounder, Developer, \etc).

The achieved results reported five libraries as the ones used by developers for network requests: (i) {\tt Retrofit}, mentioned by 65 participants; (ii) {\tt OkHttp}, 61 participants; (iii) {\tt HttpURLConnection}, 23; (iv) {\tt Volley} 20; and (v) {\tt Apache HTTP client}, with 19 participants. Since from Android 6.0 {\tt Apache HTTP} is no longer recommended and its employment has been discouraged by Google, we decided to exclude it from the final list of libraries to be supported in \approach. Concerning the rest of the libraries, these fell under a \textit{long-tail} area, thus we included only the first four listed by developers. The complete list of libraries mentioned by the surveyees is available in our replication package \cite{replication}.

\subsection{Connectivity Issues Supported  by \approach}
\label{sub:connIssuesElicit}

In the context of our work, a \emph{connectivity issue} is a suboptimal implementation choice that can affect the correct behavior of an Android app when used in \emph{zero/unreliable connectivity scenarios}. As shown in the literature, these issues can cause crashes in apps leading to user experience issues \cite{escobar:emse2021}. In this section we describe the six categories of issues we support in \approach, while the complete list of 16 issue types (belonging to these six categories) is presented in \tabref{tab:eci_summary}. The specific detection strategies are described in \secref{sub:connIssuesDetect}. 

\vspace{0.5em}
\subsubsection{Connect to network permissions (CNP) \vspace{0.5em} }~\\
\indent When performing network operations within an Android app, the manifest file must include some essential permissions: (i) \code{INTERNET}, allowing full network access to the app, thus being able to open network sockets and use custom network protocols; and (ii) \code{ACCESS\_\-NETWORK\_\-STATE}, allowing to access information about network connections such as which networks exist and to which ones the device is connected \cite{googleGuidelines}. 

Despite the fact that both permissions are \emph{normal permissions}, which means they are granted at install time and do not need to be requested at runtime, they must be explicitly included in the manifest \cite{googlePermissions}. Since the Android manifest file is manually written by developers, misconfigurations may happen, such as the lack of permissions that are actually required by the app~\cite{Jha2017}.

\vspace{0.5em}
\subsubsection{Connection Status Check (CSC)\vspace{0.5em}}~\\
\indent Before performing network operations within an app it is recommended \cite{googleGuidelines,EscobarTSE2020} to validate the state of the network connectivity, and in particular: (i) the type of network (\eg \code{WI-FI}, \code{MOBILE}, \code{ETHERNET}) that would allow the developers to check, for example, whether a \emph{Wi-Fi} network is available in case large amounts of data must be downloaded (to avoid the user to incur in unexpected expenses); (ii) its availability, as a user could be experimenting flaky mobile networks, airplane mode, and restricted background data; and (iii) its Internet capabilities, in consideration of a scenario in which a user is connected to a certain network not providing Internet access. These validations should be implemented globally in the project and before triggering a network request, thus preventing an app from inadvertently using wrong configurations. Also, if a network connection is unavailable, the application should be able to handle such a case via the implementation of \emph{offline} practices \cite{escobar:emse2021,archibald202, connForBillions}. 

\vspace{0.5em}
\subsubsection{Network usage management (NMG)\vspace{0.5em}}~\\
\indent Fine-grained control over the usage of network resources should be provided to the user if an application constantly performs network operations \cite{googleGuidelinesManaging}. The user should be able to modify such settings, \eg the frequency at which the app syncs data or whether to trigger network operations only when the device is connected to \emph{Wi-Fi}. To provide the user with such a possibility, an \code{Activity} offering options to control data usage should be defined within the Android manifest.

\vspace{0.5em}
\subsubsection{Response handling (RH)\vspace{0.5em}}~\\
\indent Libraries providing HTTP services use callback functions for response and error handling. Regarding ``successful'' \code{onResponse()} callbacks and notably when back-end services are not under the app developers control, it is critical to interpret the data provided by the external service (\eg HTTP message body data and HTTP status codes) since the lack of routines for handling exceptional scenarios can cause broken data manipulation, thus leading to a crash. Therefore, it is important to validate possible errors on the response in order to make the app react properly in such scenarios. Concerning the \code{onFailure()} callbacks, it is indispensable to employ the error callback to (i) handle unexpected behaviors and (ii) improve the user experience by presenting \emph{informative messages} aimed at properly conveying the errors \cite{escobar:emse2021, EscobarTSE2020}.

\vspace{0.5em}
\eject
\subsubsection{Task scheduling (TS)\vspace{0.5em}}~\\
\indent It is not recommended to implement network operations performing \emph{synchronous} requests. 

Indeed, the latter blocks the execution of code which may cause an unresponsive behavior from the app. Note that this is not an Android-specific issue \cite{MicroSyncASync, mozillaAsyncSync}. Avoiding \emph{synchronous} requests reduces the chance of running network operations in the main thread, something that it is discouraged and also reported with the throwing of a \code{NetworkOnMainThreadException} \cite{escobar:emse2021, googleThreads,EscobarTSE2020}.  

Complementary, to reliably schedule asynchronous tasks that are expected to be properly executed even under flaky circumstances, the \code{WorkManager} API \cite{workManager} is the recommended scheduling API in Android, and should be preferred to older managers (\eg \code{FirebaseJobDispatcher}, \code{GcmNetworkManager}, and \code{Job Scheduler}).

\vspace{0.5em}
\subsubsection{Library specific issues (LBS)\vspace{0.5em}}~\\
\indent While the above-presented categories of connectivity issues can affect network operations implemented using different libraries, we also identified a well-known anti-pattern concerning the \emph{OkHttp} library. According to its documentation \cite{okhttp}: \textit{``OkHttp performs best when you create a single OkHttpClient instance and reuse it for all of your HTTP calls. This is because each client holds its own connection pool and thread pools. Reusing connections and threads reduces latency and saves memory. Conversely, creating a client for each request wastes resources on idle pools..."}. 


\begin{table*}
	\small
	\caption{List of \eci detected by \approach and their respective coverage given a network library\vspace{-0.2cm}}
	\resizebox{\textwidth}{!} {
	\begin{tabular}{@{\extracolsep{\fill}} |c|c|c|c|c|c|c|c|c| } 
	
		\hline
		Category & ID & Issue & Retrofit & OkHttp  & Volley  & HttpURLConnection & Resource\\
		\hline
	
		\multirow{2}{4em}{CNP} 
		
		& INP & No \code{INTERNET} Permission &  \checkmark&\checkmark  & \checkmark& \checkmark & \cite{googleGuidelines,googlePermissions,Jha2017,olmstead2015apps} \\ 
		& ACP & No \code{ACCESS\_NETWORK\_STATE} Permission &  -& - & -& - &  \\ 
	
		\hline\hline
		
		\multirow{6}{4em}{CSC} 
		& NP & No Network connection check in Project  &  \checkmark&\checkmark  & \checkmark& \checkmark & \cite{googleGuidelines, escobar:emse2021,archibald202,EscobarTSE2020, hellman2014android, connForBillions}\\ 
		& NM & No Network connection check in Method &  \checkmark&\checkmark  & \checkmark& \checkmark &  \\
		& TP  & No Network type check in Project &  \checkmark&\checkmark  & \checkmark& \checkmark &  \\
		& TM & No Network type check in Method & \checkmark&\checkmark  & \checkmark& \checkmark &  \\
		& IP & No Internet availability check in Project &  \checkmark&\checkmark  & \checkmark& \checkmark &  \\
		& IM & No Internet availability check in Method & \checkmark&\checkmark  & \checkmark& \checkmark &  \\
		
		\hline\hline
		
		\multirow{1}{4em}{NMG} 
		& AMN & No \code{ACTION\_MANAGE\_NETWORK\_USAGE} &  \checkmark&\checkmark  & \checkmark& \checkmark &   \cite{googleGuidelinesManaging}\\ 
		
		\hline\hline
		
		\multirow{4}{4em}{RH} 
		&  RI & No \code{Response} implementation & \checkmark&\checkmark  & \checkmark& N/A & \cite{escobar:emse2021,EscobarTSE2020, hellman2014android, cwe389, cwe1069, cwe1071} \\ 
		& RB & No \code{Response} body verification & \checkmark&\checkmark  & \checkmark& \checkmark &  \\
		& RC & No \code{Response} code verification & \checkmark&\checkmark  & \checkmark& \checkmark &  \\
		& OF & No \code{onFailure} implementation &  \checkmark&\checkmark  & \checkmark& \checkmark &  \\
		
		\hline\hline
		
		\multirow{2}{4em}{TS} 
		& SYN & Avoid \emph{synchronous} calls &  \checkmark&\checkmark  & N/A& \checkmark & \cite{escobar:emse2021, googleThreads, workManager,EscobarTSE2020, Lingling2018} \\ 
		& WM & No \code{WorkingManager} employment &  -& - & -& - &  \\
		
		\hline\hline
		
		\multirow{1}{4em}{LBS} 
		& OK & No more than one \code{OkHttp} Constructor &  N/A& \checkmark &  N/A&  N/A &\cite{okhttp}  \\ 
		
		\hline
	\end{tabular}
}
 \label{tab:eci_summary}
\end{table*}

\subsection{ Detection Strategies used by \approach}
\label{sub:connIssuesDetect}

The majority of the issues discussed in \secref{sub:connIssuesElicit} are related to network operations triggered by the libraries used for handling network requests. Therefore, to detect potential connectivity issues it is crucial to identify in the app's code the methods generating network requests. We refer to the points in code in which a network operation is triggered as a ``network operation trigger'' (\nt). The identification of a \nt is supported by the built-in call graph and interprocedural analysis provided by Android Lint that eases the identification of method calls raised within an app. This allows to easily identify (i) the \nt, since for each library we support, we know what the API methods used to start a network operation are (see next paragraph); and (ii) specific methods from the Android APIs that allow to, \eg check connectivity, verify the internet access, \etc These will be used in some of the detection rules described in the following.

For both \emph{Retrofit} and \emph{OkHttp} the network requests can be triggered by invoking \code{enqueue()} and \code{execute()}. The former follows an \emph{asynchronous} approach whilst the latter employs a \emph{synchronous} mechanism. \emph{HttpURLConnection} opens a connection to a remote object given a URL by employing the \emph{synchronous} \code{openConnection()} method and it lacks of an \emph{asynchronous} counterpart. Oppositely, \emph{Volley} only operates \emph{asynchronously} through the \code{add()} method. The current section discusses every detection strategy we implemented to detect the \eci. 

A summary of the 16 supported \eci is presented in \tabref{tab:eci_summary}, which reports: (i) the category to which each issue type belongs (among the ones described in \secref{sub:connIssuesElicit}); (ii) the name we assigned to the connectivity issue; (iii) the libraries in which it can manifest and for which we provide detection support; and (iv) the resources (\eg articles, official Android guidelines, \etc) we used as references to define such a connectivity issue.

\noindent\textbf{1) Connect to network permissions (CNP)} 

\emph{No \code{INTERNET} Permission}: If a \nt exists within the app analyzed by \approach, our tool parses the Android manifest to iterate over the declared permissions. If no \code{android.permission.INTERNET} element is found, \approach raises a warning.
	
\emph{No \code{ACCESS\_NETWORK\_STATE} Permission}: If an app que\-ries (i) the state of network connectivity or (i) the status of a network interface, both  \code{ConnectivityManager} and \code{NetworkInfo} packages must be imported. If these imports exist, \approach parses the permissions in the manifest and raises a warning if no \code{android.permission.\-ACCESS\_NETWORK\_STATE} element is found.

\noindent\textbf{2) Connection status check (CSC)}

\emph{No Network connection check (Project \& Method)}: \approach verifies whether a \nt exists and, if this is the case, it checks if the \code{isConnected()} or \code{onAvailable()} methods are invoked at least once in the project. If such methods have been invoked at least once but they are not being invoked in a specific method including a \nt, \approach notifies of the need for invoking these methods to check the connection state before the \nt. If, instead, these methods have never been used in the project but \nt exist, \approach reports such a lack highlighting all code locations including a \nt. Note that \code{isConnected()} has been deprecated in API level 29, with the recommendation of using instead \code{onAvailable()}. However, \approach still considers both methods as a viable solution to check the network status in order to be able to support also apps using an API level lower than 29.

\emph{No Network type check (Project \& Method)}: Similarly to the above-described check for the network status, assuming at least a \nt exists in the app, \approach verifies whether a call to either \code{getType()} or \code{hasTransport()} is present in the scope of the \nt defined in the project. Both methods allow the developer to know the type of a network (\eg \code{WiFi}). If these methods have never been invoked in the whole project, \approach will notify the developer of such a lack at project-level reporting all \nt in which the network type is not verified. Instead, if a specific method using a \nt is not invoking one of those methods, a method-level warning is raised. The method \code{getType()} has been deprecated in API level 28 in favor of \code{hasTransport()}. Also in this case, we preferred to support both of them in \approach to favor compatibility with older apps.

\emph{No Internet availability check (Project \& Method)}: For each \nt, \approach checks if \code{hasCapability()} is being called within its scope with either \code{NET\_CAPABILITY\_INTERNET} or \code{NET\-\_CAPABILITY\_VALIDATED} as arguments. 

The former checks if the network has Internet capabilities while the latter coveys that Internet connectivity was successfully detected. If \code{hasCapability()} is never invoked, \approach raises a project-level warning and all the methods including \nt will be flagged. 

Otherwise, \approach only locates the methods with \nt not having the Internet availability validation. Note that there are other non-standard ways to validate the Internet access; however, those are not currently supported in \approach.

\noindent\textbf{3) Network usage management (NMG)}
	
\emph{No \code{ACTION\_MANAGE\_NETWORK\_USAGE}}: If \approach identifies at least one \nt, it verifies whether the application declares in the Android manifest an activity offering options to the user to control the app's data usage. Such an activity is identified by looking for the declaration of an intent filter for the \code{ACTION\_MANAGE\_\-NETWORK\_USAGE} action. If such an intent is not present in the Android manifest, a warning is reported to the developer.

\noindent\textbf{4) Response handling (RH)}

\emph{No \code{Response} implementation}: Three of the four libraries we support (\ie all but \emph{HttpURLConnection}) allow the implementation of a \code{Response} callback through which it is possible to check that a response has been successfully returned to the HTTP request. \approach verifies if such a response callback is implemented and extend the scope of this information by using it as a preliminary check for the detection of the following issue (\ie No \code{Response} body verification). If this is not the case, a warning is raised. The detection of this issue is not supported for apps using \emph{HttpURLConnection}.
	
\emph{No \code{Response} body verification}: Given a \nt performed through \emph{Retrofit}, \emph{OkHttp} or \emph{Volley}, \approach checks if the \code{body} of their respective \code{Response} object is being processed through a decision-making statement including a \code{null} validation. For \emph{HttpURLConnection}, such validation is being addressed with the methods \code{getResponseMessage()} and \code{getInputStream()}, since they are the ones containing the content load. 
	
\emph{No \code{Response} code verification}: Similarly to the previous check, \approach looks for HTTP status code verifications within the \code{Response} object of each supported library. 

This can be done: in \emph{Retrofit}, through the method \code{code()}; in \emph{OkHttp}, using  \code{code()} or \code{isSuccess\-ful()}; in \emph{Volley}, by invoking \code{statusCode()}; and in \emph{HttpURLConnection} through \code{getResponseCode()}. \approach verifies if these methods are invoked within a decision-making statement looking for a \code{null} validation. 
	
\emph{No \code{onFailure} implementation}: In \emph{Retrofit} and \emph{OkHttp}, the error callback is called \code{onFailure()}, while \emph{Volley} handles the errors with the \code{ErrorListener}. Lastly, \emph{HttpURLConnection} does not have callbacks, and similarly to the cases in which the developers have implemented \emph{synchronous} \nt related to the rest of the libraries, \approach verifies if they are surrounded by a \code{TRY-CATCH} clause. Additionally, the \code{CATCH} statement is expected to not be empty. \approach reports a warning if (i) the method needed for the error callback is not implemented (in the case of the first three libraries), or (ii) errors are not correctly handled in \emph{HttpURLConnection}.
	
\noindent\textbf{5) Task scheduling (TS)}

\emph{Avoid \emph{synchronous} calls}: \approach detects each occurrence of \emph{synchronous} \nt, namely invocations to the  \code{execute()} method for both \emph{Retrofit} and \emph{OkHttp}. Concerning \emph{HttpURLConnection}, \code{openConnecti\-on()} itself is \emph{synchronous}. Google used to suggest that these operations should be executed with an \emph{asynchronous} workaro\-und by using \code{AsyncTask}, which is a \code{Class} intended to enable proper use of the UI thread, widely employed for this aim before Android API level 30. Currently, it is recommended to use the standard \code{java.util.concurrent} package or the Kotlin concurrency utilities (depending on the used language) or co-routines. Since this advice is recent and eventually \code{AsyncTask} will no longer be supported, \approach pinpoints every \emph{synchronous} \nt in the app under analysis. Note that the \emph{Volley} library does not provide any \emph{synchronous} operation.

\emph{No \code{WorkingManager} employment}: \code{WorkingManager} is an API intended to ease the implementation of reliable asynchronous tasks scheduling. \approach detects in the source code files the importing of old APIs that have been replaced by \code{WorkingManager}, namely \code{FirebaseJobDispatcher}, \code{GcmNetworkManager}, and \code{JobScheduler}, reporting such an issue to the developers.

\noindent\textbf{6) Library specific issues (LBS)}

\emph{No more than one \code{OkHtpp} Constructor}: \approach analyzes the entire project looking for constructor calls of \code{OkHttp\-Client()}. In case more than one instance is defined within the application a warning is raised.

\subsection{\approach Report}

\begin{figure}
	\centering
	\includegraphics[width=0.5\textwidth]{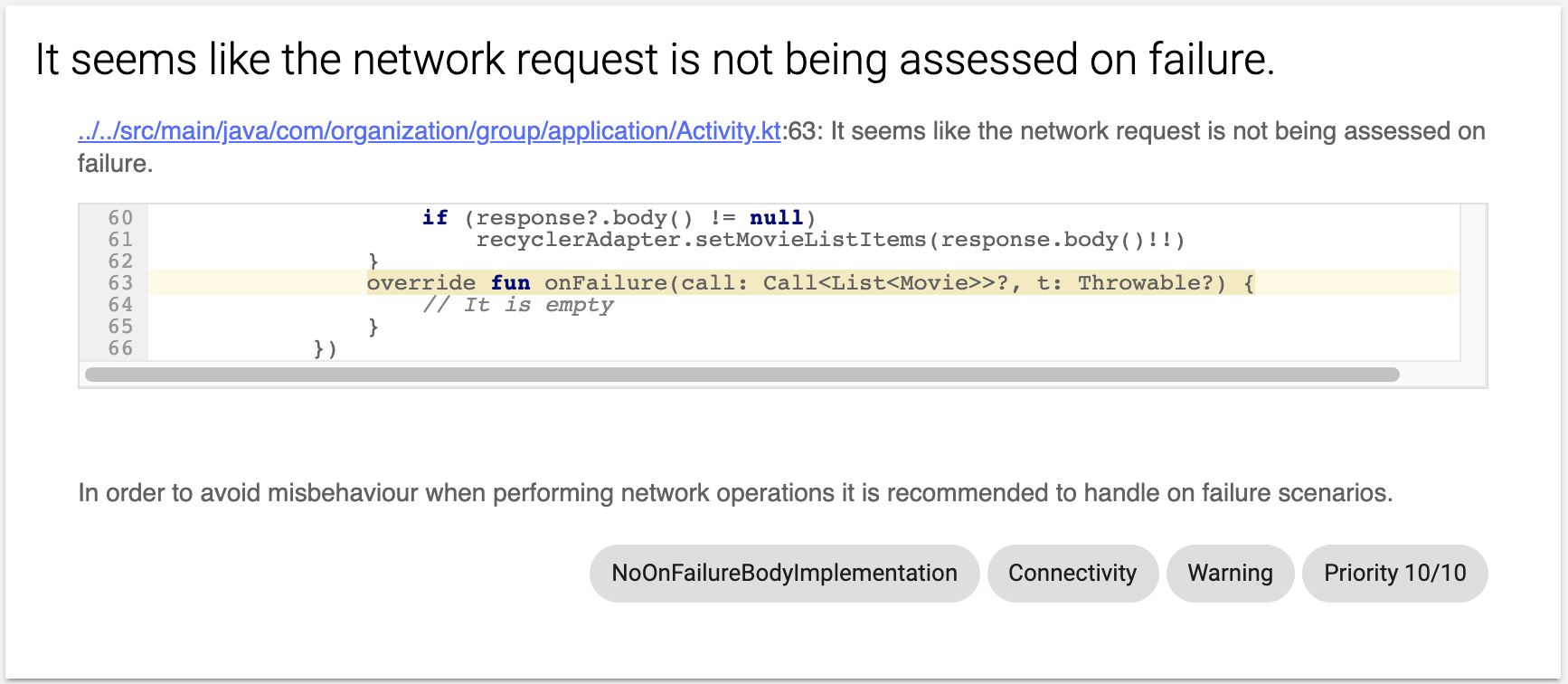}
	\captionsetup{justification=centering}
	\caption{No \code{onFailure} implementation report \vspace{-0.3cm}} 
	\label{fig:html_report}
	\vspace{-0.8em}
\end{figure}

\approach adopts the same output format of Android Lint describing the identified \eci in a structured HTML file. For each identified issue \approach provides the snippet of code in which it has been identified marking its category as \eci. Figure \ref{fig:html_report} shows an example of report for the issue type \emph{No \code{onFailure} implementation}. Android Lint, and consequently our tool, can also be used directly in the IDE. In this case, identified issues will be marked directly in the code editor.

%% file: secs/study_design.tex

\section{Study Design}
The \emph{goal} of this study is to assess (i) the precision of \approach in detecting \eci in Android apps, and (ii) the relevance of the identified issues from the practitioners' perspective. The \emph{context} is represented by (i) \numapps open source apps on which we run \approach looking for \eci;  (ii) six interviews conducted with practitioners working with Android apps; and (iii) \numissues issues we opened in \openSource open source apps to report the \eci identified by \approach.

We aim at answering the following research questions:

\textbf{RQ$_1$:} \emph{What is the precision of \approach in identifying \eci in Android apps?} We run \approach on \numapps open source apps, and we manually validated the reported \eci, classifying them as true or false positives. We discuss \approach's accuracy for the different types of issues it is designed to detect (see \secref{sec:approach}).

\textbf{RQ$_2$:} \emph{What is the relevance of the \eci identified by \approach for Android practitioners?} We answer this research question in two steps. First, we conducted six interviews with Android practitioners asking them the relevance of the \eci that \approach supports. Note that in these interviews practitioners did not inspect actual instances detected by \approach, but were rather asked for their opinion about the \emph{types} of issues we support (\tabref{tab:eci_summary}). Second, we opened \numissues issues in the issue tracker of \openSource of the \numapps apps analyzed in RQ$_1$ to collect developers' feedback about the \eci we identified. In this case, specific issue instances are studied.

\subsection{Data Collection and Analysis}
To answer our research questions, we selected \numapps Android open source apps as subject systems. As a starting point we adopted the two following datasets:

\textbf{Geiger \etal \cite{pascarella2018osprojects}.} Composed of 8,431 real-world open-source Android apps. It combines source and history information from GitHub with metadata from Google Play store.

\textbf{Coppola \etal\cite{coppola2019migrationkotlin}.} Composed of 1,232 Android apps. The authors of this dataset mined all projects hosted on F-Droid\cite{Fdroid}, a repository for free and open source Android apps.

After removing duplicates among both datasets we obtained 8,157 unique open source Android apps, from which we excluded the ones no longer available on GitHub, leaving 8,003 apps. Then, we filtered out those not written in Java and/or Kotlin, obtaining 6,847 apps. This new dataset was then analyzed to identify repositories that could support an automated running of \approach. This basically translates in the requirement that the repositories must use Gradle as their project-building tool. Thus, we identified repositories that: \circled{1} contain the source code of more than one Android app (70); \circled{2} do not use Gradle (1,902); \circled{3} use Gradle, without however adopting the default name (\ie \textit{app}) for the main module to build (3,522); and \circled{4} use Gradle, by adopting a customized name for the main module (1,353). We discarded repositories from \circled{1} and \circled{2}, since they do not support the automated running of \approach. Also, we excluded all repositories not using a version of Gradle equal or newer than 6.7.1. Indeed, accordingly to the Android Lint documentation \cite{UserGuide, ApiGuide}, the version in which we integrated \approach (30.0.0-alpha10) works properly on projects adopting those Gradle versions. 

After these filters, we obtained a sample of 62 repositories belonging to \circled{3} and 160 repositories belonging to \circled{4}. For the projects in \circled{4} (\ie not using the default name for the main module), we manually identified such a name. The final set of \numapps open-source apps was obtained as the set of projects successfully compiling and notifying at least one incident (\ie a warning raised by \approach) when running \approach on their latest snapshot (as of July 19th, 2021). On average, these apps have 1.6k commits, 36 releases, and 520 issues. 

\begin{table*}[h!]
	\centering
    \footnotesize
        \caption{Number of candidate \eci identified by \approach in the \numapps subject apps}
        \begin{tabular}{llrr|rrrr}
                \toprule
               \multirow{2}{*}{\textbf{ID}} & \multirow{2}{*}{\textbf{Issue Type}} & \textbf{\#Affected} & \textbf{\#Candidate} &  \textbf{\#Manually} & \textbf{True} & \textbf{False} & \multirow{2}{*}{\textbf{Precision}}\\
                && \textbf{Apps} & \textbf{Instances} & \textbf{Validated} & \textbf{Positives} & \textbf{Positives} & \\\midrule
        NP & No Network connection check in Project& 16 & 16 &16&16&0&100\%\\
        NM & No Network connection check in Method& 26 & 163 &50&50&0&100\%\\
        TP & No Network type check in Project& 20 & 20 &20&20&0&100\%\\
        TM & No Network type check in Method& 26 & 166 &50&50&0&100\%\\
        IP & No Internet availability check in Project& 25 & 25 &25&25&0&100\%\\
        IM & No Internet availability check in Method& 27 & 191 &50&50&0&100\%\\
        AMN & No \code{ACTION\_MANAGE\_NETWORK\_USAGE}& 42 & 42 &42&42&0&100\%\\
        RI & No \code{Response} implementation& 1 & 1 &1&0&1&0\%\\
        RB & No \code{Response} body verification& 9 & 16 &16&1&15&6.25\%\\
        RC & No \code{Response} code verification& 8 & 15 &15&6&9&40\%\\
        OF & No \code{onFailure} implementation& 13 & 45 &45&10&35&22.22\%\\
        SYN & Avoid \emph{synchronous} approach& 23 & 189 &50&41&9&82\%\\
        WM & No \code{WorkingManager} employment& 3 & 3 &3&3&0&100\%\\
        OK & No more than one \code{OkHttp} Constructor& 6 & 6&6&0&6&0\%\\\midrule
        \textbf{Total} & & & \textbf{\candidates} & \textbf{389} & \textbf{314} & \textbf{75} & \textbf{}\\\bottomrule
        \end{tabular}
    \label{tab:rq1}
\end{table*}

\vspace{1em}
\subsubsection{RQ$_1$: What is the precision of \approach in identifying Connectivity issues in Android apps?}

\approach detected \candidates candidate issues, distributed by app and type of issue as reported in the first three columns of \tabref{tab:rq1}. Note that for two \eci (\ie those belonging to the CNP category in \tabref{tab:eci_summary}) we did not find any instance in the analyzed apps, excluding them from RQ$_1$. Two authors performed a manual validation on a sample of the \candidates candidate instances. We performed a stratified sampling: For each \eci type $T_i$ we targeted the random selection of 50 of its instances by considering the analyzed apps as strata. This means that the number of $T_i$ instances we selected from each app is proportional to their prevalence in the app. For example, if $T_i$ affects three apps with 70, 20, and 10 instances, respectively, when extracting a random sample of 50 instances we picked 35, 10, and 5 instances from the three apps. 

There are two exceptions to this procedure: (i) when an app is only affected by a single $T_i$ instance, we always selected such an instance for manual inspection; (ii) if $T_i$ had less than 50 instances across all apps, we picked all its instances for manual analysis. 

Overall, we manually analyzed \manual instances, ensuring a significance interval (margin of error) of $\pm5\%$ with a confidence level of 99\% \cite{Rosner2011}. The two authors involved in the manual validation independently inspected each instance, classifying it as a true positive (TP) or false positive (FP). This was done by inspecting the source code of the apps. Once completed, the two authors together inspected the 44 cases of conflict (11.31\%), solving them through open discussion. The inspected instances together with the output of the manual validation are available in our replication package \cite{replication}.

Using the results of our manual analysis we computed the precision, \ie TP/(TP+FP), of \approach for each of the supported issue types, thus answering RQ$_1$. Note that we did not assess \approach's recall since this would require the manual analysis of the entire code base of all subject apps.

\subsubsection{RQ$_2$: What is the relevance of the \eci identified by \approach for Android practitioners?}
We answer RQ$_2$ in two steps. First, we conduct six interviews with Android practitioners in our contact network asking them about the relevance of the issue types currently supported in \approach. The first part of the interview asked participants a few demographics questions including (i) the country in which they were located; (ii) their role in the company (\eg developer, tester); (iii) their programming experience and Android programming experience (in years). Then, we asked a few questions related to whether (i) they have experienced connectivity issues as app users; (ii) they adopt connectivity strategies (\ie code aimed at ensuring the proper working on the app in different connectivity scenarios) in their apps; (iii) they think connectivity strategies are part of the prioritized tasks when developing an app. After that, we asked the participants to assess their perceived relevance of the 16 \eci that \approach is able to detect. 

For each issue type, we showed the participant a description of the issue and asked their perceived relevance of this issue on a five-point scale: \emph{Very Irrelevant}, \emph{Irrelevant}, \emph{Neutral}, \emph{Relevant}, \emph{Very Relevant}. The last question of the interview was: \textit{What kind of support would you like to have from an automated tool that helps to detect these issues?} 
We summarize in \secref{sec:results} the participants' perceived relevance of the issue types we support.

Second, using the true positive instances (\ie actual \eci) manually identified in RQ$_1$, we opened \numissues issues in \openSource of the \numapps apps involved in our study (some were excluded due to the lack of true positive instances to report). When the same issue type affected the same app in multiple locations (\eg several methods of the app used network connectivity without first checking for the status of the network), we only opened one issue summarizing all of them. For this reason, the \numissues reports concern a total of 150 \eci detected by \approach. We discuss in \secref{sec:results} the number of issue reports in which developers (i) reacted by posting a comment to our issue; (ii) confirmed the issue by just commenting; or (iii) confirmed the issue and also took proper action to fix it. We also report qualitative examples of issues confirmed and ``discarded'' by developers.


%% file: secs/results.tex

\section{Results} \label{sec:results}

We discuss the achieved results accordingly to the formulated RQs. We start by discussing the results related to RQ$_1$ (\secref{sub:rq1}) focusing on the data reported in \tabref{tab:rq1} and complementing its discussion with qualitative examples. The results achieved for RQ$_2$ are instead presented in \secref{sub:interviews} (relevance of the issue types identified by \approach as perceived by the six interviewed practitioners) and \secref{sub:issues} (perceived relevance of the issue instances identified by \approach in open source projects). We provide in our replication package~\cite{replication} (i) the \approach tool, which is available as a JAR file with instructions on how to run it; (ii) the list of \candidates \eci detected by \approach; (iii) the transcriptions of our interviews (translated from Spanish to English); (iv) the list of opened issues; and (v) the list of analyzed apps.

\subsection{RQ$_1$: Precision of \approach}
\label{sub:rq1}

Looking at \tabref{tab:rq1}, it is first important to notice the lack of specific \eci, for which no instances have been identified in our study. These are the issues belonging to the network permissions, and in particular ``No \texttt{INTERNET} permission'' and ``No \texttt{ACCESS\_NETWORK\_STATE} permission''. 

We believe that this result is a consequence of our experimental design, in which we run \approach on the latest available snapshot of the subject apps. Indeed, issues related to permissions are more likely to affect the app in the early stages of the software lifecycle, since those issues are easy to spot, not allowing the app to properly work.

Concerning the \eci within the Connection Status Check (CSC) category, this block of issues comprises most of the identified instances, with a total of 581 candidate instances, accounting for 64\% of total instances. Based on our manual validation, all the issues in the CSC category were detected with a 100\% precision (see \tabref{tab:rq1}). 

\begin{figure}[htp]
	\centering
	
	\subfloat[Lack of connection status check in \textbf{mtg-familiar}]{%
		\includegraphics[clip,width=0.85\columnwidth]{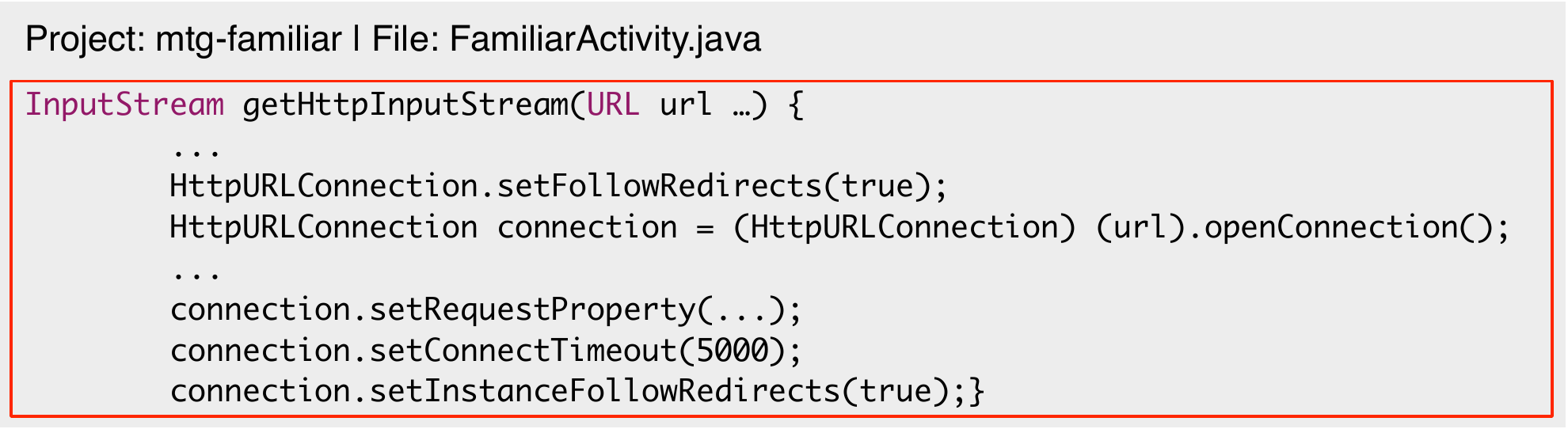}%
		\label{fig:csc-mtg-familiar}
	}
	\vspace{-0.5em}
	\subfloat[Lack of connection status check in \textbf{OpenSongTablet}]{%
		\includegraphics[clip,width=0.85\columnwidth]{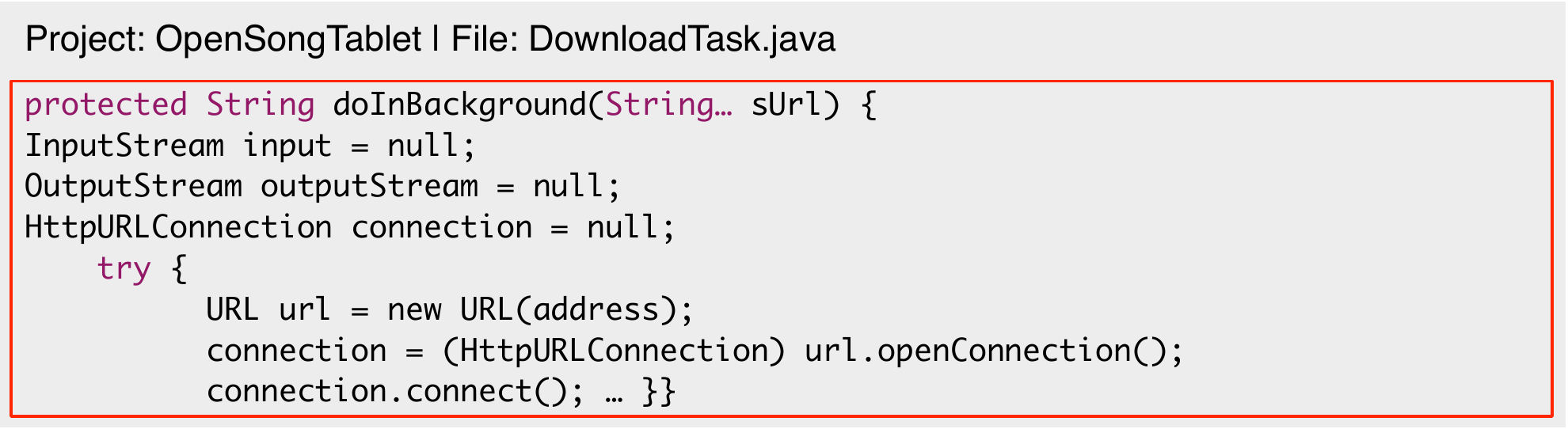}%
		\label{fig:csc-opensong}
	}
	
	\captionsetup{justification=centering}
	\caption{Missing connection status verification}
	\vspace{-0.5em}
\end{figure}

The CSC issue depicted in \figref{fig:csc-mtg-familiar} is a representative example of a network operation performed without verifying the \textit{network connection}, \textit{network type} nor the \textit{Internet availability}. The presented snippet of code is in charge of opening an \code{InputStream} to the HTML content of the given URL. The \nt in this scenario is \code{openConnection()}, which returns a \code{URLConnection} instance that represents a connection to the remote object. Similarly, in Open Song Tablet (see \figref{fig:csc-opensong}), the developers are downloading songs from a URL, also in this case implementing a \nt with \code{openConnection()}. Despite the downloading operation is performed within a \code{TRY-CATCH} block that can handle eventual exceptions, the lack of specific checks for network connection (NM), network type (TM), and Internet availability (IM) does not allow to notify the user about specific errors occurring when running the \nt, leaving the user with a generic error. The lack of informative messages in case of connectivity errors is one of the main issues recently highlighted in the literature \cite{escobar:emse2021}.

Out of the 44 inspected apps, 42 of them do not implement the intent filter \code{ACTION\_MANAGE\_NETWORK\_USAGE} (AMN issue), meaning that they do not allow their user to have fine-grained control over the app's usage of network resources. Also in this case all detected instances we manually validated were classified as true positives. Differently, for issues related to the \textit{response handling}, \approach's precision is not consistent across the different issue types. 

Concerning \textit{RI} (\ie the lack of \code{onResponse} implementation), \approach only reported a single instance that was classified as a false positive, since developers used a non-standard mechanism to handle the request. Also in this case, it is likely that this type of issue affect the app in the early development stages. In fact, in mature projects the \code{onResponse} implementation is expected. Thus, the identification of this type of issue is probably more useful when using \approach inside the IDE rather than when running it on newer versions of the app.

Similarly, the identification of \textit{RB} (no response body verification) and \textit{OF} (no \code{onFailure} implementation) also resulted in low precision values, being 6.25\% and 22.2\%, respectively. This suggests that the detection rules we defined must be revised and refined to handle special cases we did not consider. For example, in case of \textit{RB}, among the cases highlighted as false positive, we noticed that developers were assigning the \code{body()} content to a variable and then a \code{null} check was performed. However, \approach checks if the \code{body} given by a \code{Response} object is being processed by the app through a \textbf{direct} decision-making statement such as a \code{null} validation, meaning that this heuristic must be enhanced in future releases of our approach. Moreover, concerning \textit{OF}, in some false positives we found the presence of logging elements (\eg \code{Log.d()}) printing the errors in response to a failure. While these are ``borderline'' strategies to handle errors, we still decided to consider them as a false positives. A better precision (40\%) was achieved by \approach when detecting instances of \textit{RC} (no response code verification). Still, also here margins for improvement are present.

\begin{figure}[htp]
\centering
	\includegraphics[clip,width=0.85\columnwidth]{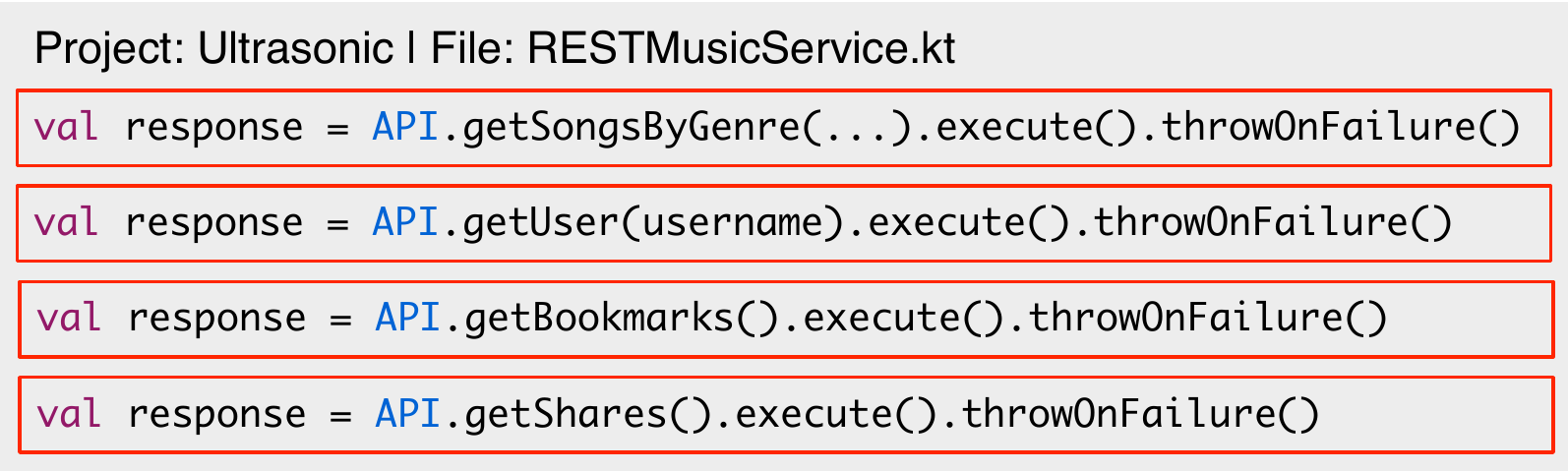}%
	\captionsetup{justification=centering}
	\caption{Several \textit{synchronous} operations in \textbf{Ultrasonic}}
		\label{fig:syn-ultra}
		\vspace{-0.3cm}
\end{figure}

Moving to the issues related to the \textit{task scheduling}, \ie ``avoid \emph{synchronous} approach'' (\textit{SYN}) and ``no \code{WorkingManager} employment'' (WM), the precision is quite high, with 82\% and 100\% of correctly detected instances, respectively. However, for the ``no \code{WorkingManager} employment'' issue we only assessed the three candidate instances \approach identified. Several examples of \textit{SYN} issues within the \textbf{Ultrasonic} app are shown in \figref{fig:syn-ultra}. Developers are using \code{OkHttp}'s \code{execute()} method in order to synchronously retrieve some resources from an external server. This issue was detected in an Kotlin file.

Lastly, for the issue specific to \code{OkHttp}, \ie no more than one \code{OkHttp} Constructor (\textit{OK}), \approach found six instances all labeled as false positives. 

This is because \approach was run not only on the source code of the subject apps, but also on the code of their dependencies. This means that it identified several instances of \code{OkHttp} clients that, however, were not due to multiple instances declared within the app, but in the app and in its dependencies. 

This can be avoided by simply setting an Android Lint property, indicating that it must only be run on the app's code. Note also that such a setting only affected the detection of this specific type of issue (\ie all other reported instances were related to the app's source code).%
\input{secs/interviews}
\input{secs/openissues}

%% file: secs/interviews.tex
\subsection{RQ$_2$: Relevance of connectivity issues (Interviews)}
\label{sub:interviews}

\begin{figure}[!ht]
	\centering
	\includegraphics[width=1\linewidth]{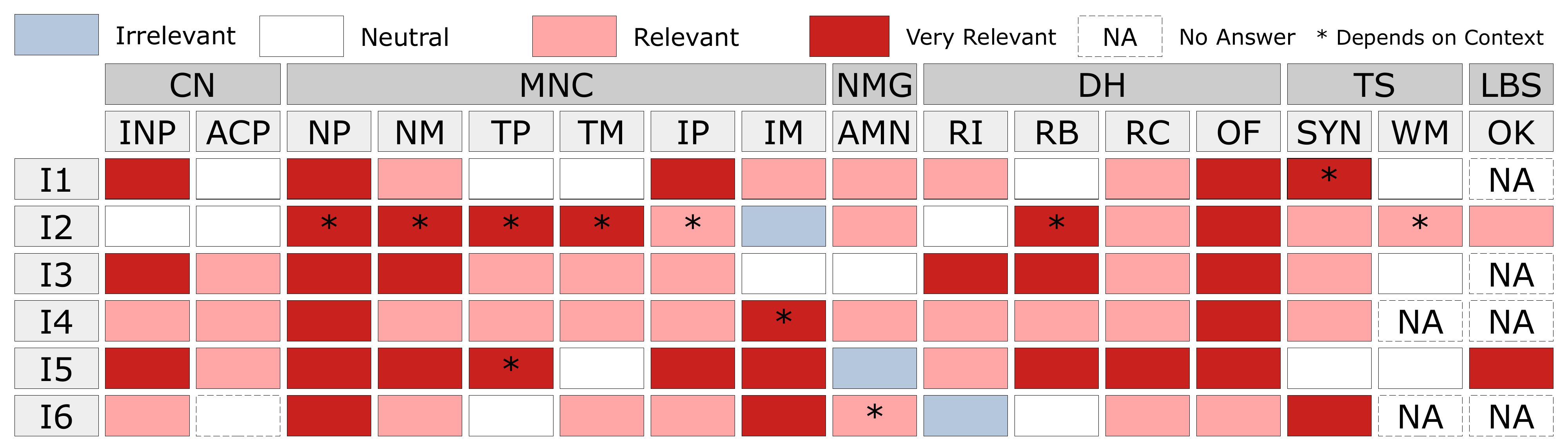}
	\captionsetup{justification=centering}
	\caption{Issue relevance classification from interviews} 
	\label{fig:issue_relevance}
	\vspace{-0.3cm}
\end{figure}

We focus our discussion on the relevance of the types of \eci \approach supports as perceived by the six interviewed Android practitioners (\fref{fig:issue_relevance}), while the answers they provided to all our questions are available in the replication package \cite{replication}.  Participants have, on average, $\sim$7 years of overall programming experience and approximately $\sim$3 years concerning Android engineering tasks. The rows in the heat map represent the six interviewed practitioners (indicated with I$n$, with $n$ going from 1 to 6), while the columns are the types of issues we investigated. The rectangle at the intersection participant/issue-type can have different colors based on the specific answer we got. First, we use a 5-step color scale from blue to red representing the reported relevance: dark blue for \textit{very irrelevant}, light blue for \textit{irrelevant}, white  for \textit{neutral}, light red for \textit{relevant}, and dark red for \textit{very relevant}. Additionally, there are 2 marks that are used in the figure to represent: (i) the cases in which a participant did not answer the question (\ie a \textbf{NA} is shown in the rectangle); and (ii) cases in which the interviewee stated that the relevance of the issue depends on the app analyzed (\eg by how important is connectivity for that app)---an asterisk inside the rectangle is shown in this case. In the latter case, \fref{fig:issue_relevance} reports the highest relevance perceived by the interviewee (\eg the answer ``it can be very relevant but it depends on the connectivity usage of the app'' is represented with a dark red rectangle with an asterisk inside).

As it can be seen from \fref{fig:issue_relevance}, none of the issues was tagged as \textit{very irrelevant} by the interviewees. Nevertheless, three of the interviewees reported one case (each) as irrelevant. Specifically, I2 said that checking Internet availability at method level is irrelevant if there is already an existent check at project level; I5 claimed that allowing users to manage network usage is irrelevant since efficient design processes should contemplate that factor; finally, I6 said that checking the implementation of response handling is irrelevant since they never saw such a problem in any of the app they worked on. 

However, as it can be seen from \fref{fig:issue_relevance}, for each of these three issues there were other developers finding them relevant.

The issues perceived as more relevant by practitioners are: (i) \emph{No Network connection check in Project}, (ii) \emph{No onFailure implementation}, (iii) \emph{No Network connection check in Method}, and (iv) \emph{No Internet availability check in Project}. Basically, issues related to the lack of checks for connectivity and the lack of handling error conditions. For example, I1 talking about the \emph{No Network connection check in Project} issue, said:  ``\textit{In our case, it would be very relevant. In our apps we start by validating that a network connection exists; we have a background process for that that starts with the opening of the application, it is like a daemon continuously checking}''.

I5 talking about the \emph{No onFailure implementation} issue said:

``\textit{Yes, I have seen it and I think it is very relevant. In fact, for me it is so relevant that I usually prefer to implement the error paths and then the success path, because in a distributed environment I do not know what is happening to the device}''. 

On the other side there are issue types that practitioners do not find relevant. This is the case of the \textit{No \code{ACCESS\_NETWORK\_STATE} Permission} (ACP) that was marked by three interviewees as ``neutral'' in terms of relevance, since they claimed that the associated permission is added along with the \textit{\code{INTERNET} Permission}. Also, interviewees do not find the \emph{No \code{WorkingManager} employment} (WM) as an actual issue, since the benefits provided by this component for asynchronous tasks are considered minor. Finally, four of the six participants do not use the \textit{OkHttp} library and, thus, did not assess the relevance of the \emph{No more than one \code{OkHttp} Constructor} (OK), that was marked as relevant and very relevant by the remaining two practitioners.

As a final note, it is worth noticing that five out of the six participants indicated that the relevance of at least one of the assessed issue types depend on the specific context (\ie app) in which it manifests and, more in particular, on the importance that connectivity has in the affected app.

To summarize, we found that: (i) for 13 out of the 16 issue types, there is a majority of participants (at least 4 out of 6) assessing them as \emph{relevant} or \emph{very relevant}; and (ii) the relevance of specific issues is influenced by the context in which they appear.

%% file: secs/openissues.tex
\subsection{RQ$_2$: Relevance of connectivity issues (OSS projects)}
\label{sub:issues}
 
We opened \numissues GitHub issues in 27 of the \numapps apps we studied. These issues represent a total of 150 \eci since, as previously said, a single opened issue can concern multiple instances of the same issue type (\eg multiple instances of \emph{No Network connection check in Method}). Once opened the issues, in the following six weeks we interacted with the developers who reacted by posting a comment and/or taking other types of action. After the six weeks, we summarized the status of the \numissues issues.

Out of \numissues, 53 resulted in a reaction from the developers, which means that either they provided us a feedback by commenting or they modified the status of the issue (\eg by closing it, adding a tag). 

Out of these 53 issues, 15 have been recognized as \eci, and for 8 of them the developers implemented a change to address the issue. For example, developers of \textit{OpenSongTablet} commented: ``\textit{Many thanks for alerting me to this. This issue is addressed in an upcoming version that is still in early alpha [...]}''. 

Concerning the remaining 38 issues, in 14 cases the issue was just closed without any interaction with the developer. The remaining ones were tagged by the developers as false issues, leading to the their closure. Only on 24 of those the developers provided us feedback regarding the issues. In one of those, developers pointed to the low relevance of the type of issue (\ie \emph{No Network connection check in Method}) in \textit{BinaryEye}, explaining that ``\textit{If this app would make many requests at once, then I would probably check the network connection before making all these requests. But for just one request per scan, this would be overkill}''. In the other cases, the negative outcome (\ie the lack of fixing) was due to the fact that the developers did not recognize the reported issue as a potential point of failure in connectivity scenarios.

Interestingly, we did not observe any specific pattern in terms of types of issue that were considered relevant/not relevant by developers. In other words, some issue types were considered relevant by some developers (with a consequent fixing) and irrelevant by others (with a closing of the issue).

These findings point to three observations. First, as already observed in our interviews, the relevance of the \eci we identify is substantially impacted by the context in which they appear (\ie by the affected app). Second, it is possible that some of the issues identified by \approach, while recognized in the literature or in the technical documentation, are unlikely to manifest in failures, thus being mostly considered as ``good practices'' rather than potential ``bugs'' to fix. This implies that a better reporting of the severity of the identified issue might be needed in \approach, possibly exploiting an ``adaptive'' mechanism that can change the severity level depending on the app under analysis. Third, while it is clear that more work is needed on \approach, we found as a positive result the identification of 26 confirmed \eci (reported in the 15 confirmed issues) in open source apps.

%% file: secs/threats.tex

\section{Threats to Validity}

\textbf{Construct Validity.} To assess the precision of \approach, we relied on a manual analysis independently performed by two authors, reporting the number of conflicts raised in this process. Despite such a double-check, we are aware that imprecisions are still possible, and this is one of the reasons why we share in the replication package \cite{replication} the output of our manual validation. Concerning our second research question, we acknowledge that both in the interviews and in the study involving the opening of issues, the reported results represent the subjective perception of practitioners.


\textbf{Internal Validity.} The set of \eci that \approach supports was derived by the analysis of the literature,  the official Android developer guidelines, and the documentation of the libraries used to establish HTTP connections.

\noindent We acknowledge that our contribution is limited by the fact that \approach is tied to specific versions of certain libraries and that other, and possibly more relevant \eci, might have been missed in our study. Similarly, the detection strategies we defined \eci were based on our intuition. Further research is needed to optimize/improve them.

\textbf{External Validity.} In RQ$_1$ the assessment of \approach's precision is based on a set of \manual manually validated instances. Additional instances should be inspected for better generalization as we acknowledge that static analysis approaches are prone to introduce false positives. For RQ$_2$, we preferred interviews with a limited number of practitioners rather than surveys with more participants to better understand and interpret the practitioners' perceptions of the \eci. Finally, the number of issues we opened, while limited (\numissues), still covered a total of 150 \eci.

%% file: secs/conclusion.tex

\section{Conclusion \& Future Work}

We presented \approach, a tool designed to statically detect \eci affecting Android apps. The list of 16 \eci supported by \approach has been derived from the existing literature and by studying the technical documentation of Android and of libraries usually adopted to handle HTTP connections in Android apps. The precision of the detection strategies we implemented has been evaluated by manually inspecting the candidate instances identified by \approach, showing that improvements are needed for the detection of 5 of the 16 issue types, while excellent precision is achieved in the remaining 11 cases.

We then assessed the relevance of the issue types detected by \approach by interviewing six practitioners, showing that they found most of the issues relevant, despite reporting their relevance as influenced by the specific context (app) in which they appear (\eg domain category). Such a finding has also been confirmed in the subsequent evaluation performed by opening \numissues issues in open source projects, with contradicting observations provided by the developers who managed those issues. In 15 cases, they confirmed the issue (for a total of 26 actual code issues identified, since one issue report could concern multiple \eci), while in 24 cases they closed the issue as non-relevant for their app. 

The achieved results are not fully positive nor negative, indicating the need for additional research on connectivity issues and on effective detection tools considering different Android artifacts (\eg Android Packages - APKs) and languages (\eg Flutter/DART). Also, additional studies with developers are needed to better understand the contexts in which connectivity issues are actually considered relevant.

%% file: main.bbl
\begin{thebibliography}{10}
\providecommand{\url}[1]{#1}
\csname url@samestyle\endcsname
\providecommand{\newblock}{\relax}
\providecommand{\bibinfo}[2]{#2}
\providecommand{\BIBentrySTDinterwordspacing}{\spaceskip=0pt\relax}
\providecommand{\BIBentryALTinterwordstretchfactor}{4}
\providecommand{\BIBentryALTinterwordspacing}{\spaceskip=\fontdimen2\font plus
\BIBentryALTinterwordstretchfactor\fontdimen3\font minus
  \fontdimen4\font\relax}
\providecommand{\BIBforeignlanguage}[2]{{%
\expandafter\ifx\csname l@#1\endcsname\relax
\typeout{** WARNING: IEEEtran.bst: No hyphenation pattern has been}%
\typeout{** loaded for the language `#1'. Using the pattern for}%
\typeout{** the default language instead.}%
\else
\language=\csname l@#1\endcsname
\fi
#2}}
\providecommand{\BIBdecl}{\relax}
\BIBdecl

\bibitem{Nagappan:saner2016}
M.~Nagappan and E.~Shihab, ``Future trends in software engineering research for
  mobile apps,'' in \emph{2016 IEEE 23rd International Conference on Software
  Analysis, Evolution, and Reengineering (SANER)}, vol.~5, 2016, pp. 21--32.

\bibitem{Martin:tse2017}
W.~Martin, F.~Sarro, Y.~Jia, Y.~Zhang, and M.~Harman, ``A survey of app store
  analysis for software engineering,'' \emph{IEEE Transactions on Software
  Engineering}, vol.~43, no.~9, pp. 817--847, 2017.

\bibitem{Bavota:tse2015}
G.~Bavota, M.~Linares-V\'asquez, C.~E. Bernal-C\'ardenas, M.~D. Penta,
  R.~Oliveto, and D.~Poshyvanyk, ``The impact of api change- and
  fault-proneness on the user ratings of android apps,'' \emph{IEEE
  Transactions on Software Engineering}, vol.~41, no.~4, pp. 384--407, 2015.

\bibitem{Linares:FSE13}
M.~Linares-V\'{a}squez, G.~Bavota, C.~Bernal-C\'{a}rdenas, M.~Di~Penta,
  R.~Oliveto, and D.~Poshyvanyk, ``Api change and fault proneness: A threat to
  the success of android apps,'' in \emph{Proceedings of the 2013 9th Joint
  Meeting on Foundations of Software Engineering}, ser. ESEC/FSE 2013, 2013.

\bibitem{Tian:icsme2015}
Y.~Tian, M.~Nagappan, D.~Lo, and A.~E. Hassan, ``What are the characteristics
  of high-rated apps? a case study on free android applications,'' in
  \emph{2015 IEEE International Conference on Software Maintenance and
  Evolution (ICSME)}, 2015, pp. 301--310.

\bibitem{Noei:emse2017}
E.~Noei, M.~D. Syer, Y.~Zou, A.~E. Hassan, and I.~Keivanloo, ``A study of the
  relation of mobile device attributes with the user-perceived quality of
  android apps,'' \emph{Empirical Softw. Engg.}, 2017.

\bibitem{bakar:permissions}
N.~S. A.~A. Bakar and I.~Mahmud, ``Empirical analysis of android apps
  permissions,'' in \emph{2013 International Conference on Advanced Computer
  Science Applications and Technologies}, 2013, pp. 406--411.

\bibitem{Frank2012}
M.~{Frank}, B.~{Dong}, A.~{Porter Felt}, and D.~{Song}, ``Mining permission
  request patterns from android and facebook applications,'' in \emph{2012 IEEE
  12th International Conference on Data Mining}, 2012, pp. 870--875.

\bibitem{Mostafa2017}
S.~{Mostafa}, R.~{Rodriguez}, and X.~{Wang}, ``Netdroid: Summarizing network
  behavior of android apps for network code maintenance,'' in \emph{2017
  IEEE/ACM 25th International Conference on Program Comprehension (ICPC)},
  2017, pp. 165--175.

\bibitem{olmstead2015apps}
K.~Olmstead and M.~Atkinson, ``Apps permissions in the google play store,''
  2015.

\bibitem{escobar:emse2021}
C.~Escobar-Vel{\'a}squez, A.~Mazuera-Rozo, C.~Bedoya, M.~Osorio-Ria{\~n}o,
  M.~Linares-V{\'a}squez, and G.~Bavota, ``Studying eventual connectivity
  issues in android apps,'' \emph{Empirical Software Engineering}, 2022.

\bibitem{googleGuidelines}
\BIBentryALTinterwordspacing
Google, ``Connect to the network.'' [Online]. Available:
  \url{https://developer.android.com/training/basics/network-ops/connecting}
\BIBentrySTDinterwordspacing

\bibitem{googlePermissions}
\BIBentryALTinterwordspacing
------, ``Permissions on android.'' [Online]. Available:
  \url{https://developer.android.com/guide/topics/permissions/overview}
\BIBentrySTDinterwordspacing

\bibitem{Jha2017}
A.~K. Jha, S.~Lee, and W.~J. Lee, ``Developer mistakes in writing android
  manifests: An empirical study of configuration errors,'' in \emph{2017
  IEEE/ACM 14th International Conference on Mining Software Repositories
  (MSR)}, 2017, pp. 25--36.

\bibitem{archibald202}
\BIBentryALTinterwordspacing
J.~Archibald, ``The offline cookbook.'' [Online]. Available:
  \url{https://web.dev/offline-cookbook/}
\BIBentrySTDinterwordspacing

\bibitem{EscobarTSE2020}
C.~Escobar-Vel{\'a}squez, M.~Linares-V{\'a}squez, G.~Bavota, M.~Tufano, K.~P.
  Moran, M.~Di~Penta, C.~Vendome, C.~Bernal-C{\'a}rdenas, and D.~Poshyvanyk,
  ``Enabling mutant generation for open- and closed-source android apps,''
  \emph{IEEE Transactions on Software Engineering}, 2020.

\bibitem{googleThreads}
\BIBentryALTinterwordspacing
Google, ``Processes and threads overview.'' [Online]. Available:
  \url{https://developer.android.com/guide/components/processes-and-threads}
\BIBentrySTDinterwordspacing

\bibitem{workManager}
\BIBentryALTinterwordspacing
------, ``Schedule tasks with workmanager.'' [Online]. Available:
  \url{https://developer.android.com/topic/libraries/architecture/workmanager}
\BIBentrySTDinterwordspacing

\bibitem{hellman2014android}
E.~Hellman, \emph{Android Programming : Pushing the Limits}.\hskip 1em plus
  0.5em minus 0.4em\relax Chichester, West Sussex, United Kingdom: John Wiley
  \& Sons Ltd, 2014.

\bibitem{connForBillions}
\BIBentryALTinterwordspacing
Google, ``Connectivity for billions.'' [Online]. Available:
  \url{https://developer.android.com/docs/quality-guidelines/build-for-billions/connectivity#network-offline}
\BIBentrySTDinterwordspacing

\bibitem{androidDevGuide}
``Android development guides. \url{https://developer.android.com/guide}.''

\bibitem{okhttp}
``Okhttp. \url{https://square.github.io/okhttp/}.''

\bibitem{retrofit}
``Retrofit. \url{https://square.github.io/retrofit/}.''

\bibitem{volley}
``Volley. \url{https://developer.android.com/training/volley/index.html}.''

\bibitem{HttpURLConnection}
``Httpurlconnection.
  \url{https://developer.android.com/reference/java/net/HttpURLConnection}.''

\bibitem{replication}
\BIBentryALTinterwordspacing
A.~Mazuera-Rozo, C.~Escobar-Vel{\'a}squez, J.~Espitia-Acero,
  M.~Linares-V{\'a}squez, and G.~Bavota, ``Replication package.'' [Online].
  Available: \url{https://sites.google.com/view/conan-ecn}
\BIBentrySTDinterwordspacing

\bibitem{LI201767}
L.~Li, T.~F. Bissyand{\'e}, M.~Papadakis, S.~Rasthofer, A.~Bartel, D.~Octeau,
  J.~Klein, and L.~Traon, ``Static analysis of android apps: A systematic
  literature review,'' \emph{Information and Software Technology}, 2017.

\bibitem{ParsonsM08}
T.~Parsons and J.~Murphy, ``Detecting performance antipatterns in component
  based enterprise systems,'' \emph{Journal of Object Technology}, 2008.

\bibitem{TrubianiBHAK18}
C.~Trubiani, A.~Bran, A.~van Hoorn, A.~Avritzer, and H.~Knoche, ``Exploiting
  load testing and profiling for performance antipattern detection,''
  \emph{Information {\&} Software Technology}, 2018.

\bibitem{GrechanikFuXie2012}
M.~Grechanik, C.~Fu, and Q.~Xie, ``{Automatically Finding Performance Problems
  with Feedback-directed Learning Software Testing},'' in \emph{International
  Conference on Software Engineering (ICSE)}, 2012.

\bibitem{liu2014characterizing}
Y.~Liu, C.~Xu, and S.-C. Cheung, ``Characterizing and detecting performance
  bugs for smartphone applications,'' in \emph{Proceedings of the 36th
  International Conference on Software Engineering}, 2014.

\bibitem{Sadeghi:ICSE15}
A.~Sadeghi, H.~Bagheri, and S.~Malek, ``Analysis of android inter-app security
  vulnerabilities using covert,'' in \emph{ICSE'15}, 2015.

\bibitem{Bagheri:TSE15}
H.~Bagheri, A.~Sadeghi, J.~Garcia, and S.~Malek, ``Covert: Compositional
  analysis of android inter-app permission leakage,'' \emph{IEEE Transactions
  on Software Engineering}, 2015.

\bibitem{Ahmad:MSR16}
W.~Ahmad, C.~K\"{a}stner, J.~Sunshine, and J.~Aldrich, ``Inter-app
  communication in android: Developer challenges,'' in \emph{Proceedings of the
  13th International Conference on Mining Software Repositories}, ser. MSR '16,
  2016.

\bibitem{Sadeghi:TSE16}
A.~Sadeghi, H.~Bagheri, J.~Garcia, and S.~Malek, ``A taxonomy and qualitative
  comparison of program analysis techniques for security assessment of android
  software,'' \emph{IEEE Transactions on Software Engineering}, 2016.

\bibitem{bello2019}
L.~{Bello-Jim{\'e}nez}, A.~{Mazuera-Rozo}, M.~{Linares-V{\'a}squez}, and
  G.~{Bavota}, ``Opia: A tool for on-device testing of vulnerabilities in
  android applications,'' in \emph{2019 IEEE International Conference on
  Software Maintenance and Evolution (ICSME)}, 2019.

\bibitem{Yang2013}
S.~{Yang}, D.~{Yan}, and A.~{Rountev}, ``Testing for poor responsiveness in
  android applications,'' in \emph{2013 1st International Workshop on the
  Engineering of Mobile-Enabled Systems (MOBS)}, 2013.

\bibitem{Xiong2018}
W.~{Xiong}, S.~{Chen}, Y.~{Zhang}, M.~{Xia}, and Z.~{Qi}, ``Reproducible
  interference-aware mobile testing,'' in \emph{2018 IEEE International
  Conference on Software Maintenance and Evolution (ICSME)}, 2018.

\bibitem{Zhao2019}
W.~{Zhao}, Z.~{Ding}, M.~{Xia}, and Z.~{Qi}, ``Systematically testing and
  diagnosing responsiveness for android apps,'' in \emph{2019 IEEE
  International Conference on Software Maintenance and Evolution (ICSME)},
  2019.

\bibitem{Linares:MSR14}
\BIBentryALTinterwordspacing
M.~Linares-V\'{a}squez, G.~Bavota, C.~Bernal-C\'{a}rdenas, R.~Oliveto,
  M.~Di~Penta, and D.~Poshyvanyk, ``Mining energy-greedy api usage patterns in
  android apps: An empirical study,'' in \emph{Proceedings of the 11th Working
  Conference on Mining Software Repositories}, ser. MSR 2014.\hskip 1em plus
  0.5em minus 0.4em\relax New York, NY, USA: Association for Computing
  Machinery, 2014, pp. 2--11. [Online]. Available:
  \url{https://doi.org/10.1145/2597073.2597085}
\BIBentrySTDinterwordspacing

\bibitem{Linares:TOSEM18}
\BIBentryALTinterwordspacing
M.~Linares-V\'{a}squez, G.~Bavota, C.~Bernal-C\'{a}rdenas, M.~D. Penta,
  R.~Oliveto, and D.~Poshyvanyk, ``Multi-objective optimization of energy
  consumption of guis in android apps,'' \emph{ACM Trans. Softw. Eng.
  Methodol.}, vol.~27, no.~3, Sep. 2018. [Online]. Available:
  \url{https://doi.org/10.1145/3241742}
\BIBentrySTDinterwordspacing

\bibitem{Li2013ISSTA}
D.~Li, S.~Hao, W.~G.~J. Halfond, and R.~Govindan, ``Calculating source line
  level energy information for android applications,'' ser. ISSTA 2013, 2013.

\bibitem{Behrouz2015}
R.~J. Behrouz, A.~Sadeghi, J.~Garcia, S.~Malek, and P.~Ammann, ``Ecodroid: An
  approach for energy-based ranking of android apps,'' in \emph{2015 IEEE/ACM
  4th International Workshop on Green and Sustainable Software}, 2015.

\bibitem{Hao2013}
S.~Hao, D.~Li, W.~G.~J. Halfond, and R.~Govindan, ``Estimating mobile
  application energy consumption using program analysis,'' in \emph{2013 35th
  International Conference on Software Engineering (ICSE)}, 2013.

\bibitem{Nayam2016}
W.~Nayam, A.~Laolee, L.~Charoenwatana, and K.~Sripanidkulchai, ``An analysis of
  mobile application network behavior,'' in \emph{Proceedings of the 12th Asian
  Internet Engineering Conference}, 2016.

\bibitem{Conti2018}
M.~{Conti}, Q.~Q. {Li}, A.~{Maragno}, and R.~{Spolaor}, ``The dark
  side(-channel) of mobile devices: A survey on network traffic analysis,''
  \emph{IEEE Communications Surveys Tutorials}, 2018.

\bibitem{Rapoport2017}
M.~{Rapoport}, P.~{Suter}, E.~{Wittern}, O.~{Lhotak}, and J.~{Dolby}, ``Who you
  gonna call? analyzing web requests in android applications,'' in \emph{2017
  IEEE/ACM 14th International Conference on Mining Software Repositories
  (MSR)}, 2017.

\bibitem{Cheng2017}
Z.~{Cheng}, X.~{Chen}, Y.~{Zhang}, S.~{Li}, and Y.~{Sang}, ``Detecting
  information theft based on mobile network flows for android users,'' in
  \emph{2017 International Conference on Networking, Architecture, and Storage
  (NAS)}, 2017.

\bibitem{continella17}
A.~Continella, Y.~Fratantonio, M.~Lindorfer, A.~Puccetti, A.~Zand, C.~Kruegel,
  and G.~Vigna, ``{Obfuscation-Resilient Privacy Leak Detection for Mobile Apps
  Through Differential Analysis},'' in \emph{Proceedings of the ISOC Network
  and Distributed System Security Symposium (NDSS)}, 2017.

\bibitem{Huang2019}
X.~{Huang}, A.~{Zhou}, P.~{Jia}, L.~{Liu}, and L.~{Liu}, ``Fuzzing the android
  applications with http/https network data,'' \emph{IEEE Access}, 2019.

\bibitem{Gadient2020}
P.~{Gadient}, M.~{Ghafari}, M.~{Tarnutzer}, and O.~{Nierstrasz}, ``Web apis in
  android through the lens of security,'' in \emph{2020 IEEE 27th International
  Conference on Software Analysis, Evolution and Reengineering (SANER)}, 2020.

\bibitem{Shabtai2011}
A.~Shabtai, U.~Kanonov, Y.~Elovici, C.~Glezer, and Y.~Weiss,
  ``{\textquotedblleft}andromaly{\textquotedblright}: a behavioral malware
  detection framework for android devices,'' \emph{Journal of Intelligent
  Information Systems}, 2011.

\bibitem{Crussell2014}
J.~Crussell, R.~Stevens, and H.~Chen, ``Madfraud: Investigating ad fraud in
  android applications,'' in \emph{Proceedings of the 12th Annual International
  Conference on Mobile Systems, Applications, and Services}, ser. MobiSys '14,
  2014.

\bibitem{Wei2012a}
T.~{Wei}, C.~{Mao}, A.~B. {Jeng}, H.~{Lee}, H.~{Wang}, and D.~{Wu}, ``Android
  malware detection via a latent network behavior analysis,'' in \emph{2012
  IEEE 11th International Conference on Trust, Security and Privacy in
  Computing and Communications}, 2012.

\bibitem{Moran:ICSE17}
K.~Moran, M.~Linares-Vasquez, C.~Bernal-Cardenas, C.~Vendome, and
  D.~Poshyvanyk, ``Crashscope: A practical tool for automated testing of
  android applications,'' in \emph{2017 IEEE/ACM 39th International Conference
  on Software Engineering Companion (ICSE-C)}, 2017.

\bibitem{Moran:ICST16}
K.~Moran, M.~Linares-V{\'a}squez, C.~Bernal-C{\'a}rdenas, C.~Vendome, and
  D.~Poshyvanyk, ``Automatically discovering, reporting and reproducing android
  application crashes,'' in \emph{2016 IEEE International Conference on
  Software Testing, Verification and Validation (ICST)}, 2016.

\bibitem{Adamsen:ISSTA15}
C.~Q. Adamsen, G.~Mezzetti, and A.~M{\o}ller, ``Systematic execution of android
  test suites in adverse conditions,'' in \emph{Proceedings of the 2015
  International Symposium on Software Testing and Analysis}, ser. ISSTA 2015,
  2015.

\bibitem{Azim14}
M.~T. Azim, I.~Neamtiu, and L.~M. Marvel, ``Towards self-healing smartphone
  software via automated patching,'' ser. ASE '14, 2014.

\bibitem{Liang2014}
C.-J.~M. Liang, N.~D. Lane, N.~Brouwers, L.~Zhang, B.~F. Karlsson, H.~Liu,
  Y.~Liu, J.~Tang, X.~Shan, R.~Chandra, and F.~Zhao, ``Caiipa: Automated
  large-scale mobile app testing through contextual fuzzing,'' in
  \emph{Proceedings of the 20th Annual International Conference on Mobile
  Computing and Networking}, ser. MobiCom '14, 2014.

\bibitem{MONKEY}
Google, ``Monkey,'' \url{https://developer.android.com/studio/test/monkey}.

\bibitem{Panizo2019}
L.~Panizo, A.~D{\'{\i}}az, and B.~Garc{\'{\i}}a, ``Model-based testing of apps
  in real network scenarios,'' \emph{International Journal on Software Tools
  for Technology Transfer}, 2019.

\bibitem{AndroidLint}
\BIBentryALTinterwordspacing
Google, ``Improve your code with lint checks.'' [Online]. Available:
  \url{https://developer.android.com/studio/write/lint}
\BIBentrySTDinterwordspacing

\bibitem{lintIssueIndex}
------, ``Lint issue index,''
  \url{http://googlesamples.github.io/android-custom-lint-rules/checks/index.md.html}.

\bibitem{GradleWrapper}
Gradle, ``The gradle wrapper,''
  \url{https://docs.gradle.org/current/userguide/gradle\_wrapper.html}.

\bibitem{AndroidDevelopers}
Google, ``Android developers,''
  \url{https://groups.google.com/g/android-developers}.

\bibitem{googleGuidelinesManaging}
\BIBentryALTinterwordspacing
------, ``Manage network usage.'' [Online]. Available:
  \url{https://developer.android.com/training/basics/network-ops/managing}
\BIBentrySTDinterwordspacing

\bibitem{MicroSyncASync}
\BIBentryALTinterwordspacing
Microsoft, ``Synchronous i/o antipattern - performance antipatterns for cloud
  apps.'' [Online]. Available:
  \url{https://docs.microsoft.com/en-us/azure/architecture/antipatterns/synchronous-io/}
\BIBentrySTDinterwordspacing

\bibitem{mozillaAsyncSync}
\BIBentryALTinterwordspacing
Mozilla, ``Synchronous and asynchronous requests - web apis: Mdn.'' [Online].
  Available:
  \url{https://developer.mozilla.org/en-US/docs/Web/API/XMLHttpRequest/Synchronous_and_Asynchronous_Requests}
\BIBentrySTDinterwordspacing

\bibitem{cwe389}
\BIBentryALTinterwordspacing
MITRE, ``Common weakness enumeration - error conditions, return values, status
  codes.'' [Online]. Available:
  \url{https://cwe.mitre.org/data/definitions/389.html}
\BIBentrySTDinterwordspacing

\bibitem{cwe1069}
\BIBentryALTinterwordspacing
------, ``Common weakness enumeration - empty exception block.'' [Online].
  Available: \url{https://cwe.mitre.org/data/definitions/1069.html}
\BIBentrySTDinterwordspacing

\bibitem{cwe1071}
\BIBentryALTinterwordspacing
------, ``Common weakness enumeration - empty code block.'' [Online].
  Available: \url{https://cwe.mitre.org/data/definitions/1071.html}
\BIBentrySTDinterwordspacing

\bibitem{Lingling2018}
L.~Fan, T.~Su, S.~Chen, G.~Meng, Y.~Liu, L.~Xu, and G.~Pu, ``Efficiently
  manifesting asynchronous programming errors in android apps,'' in
  \emph{Proceedings of the 33rd ACM/IEEE International Conference on Automated
  Software Engineering}, ser. ASE 2018, 2018.

\bibitem{pascarella2018osprojects}
F.-X. Geiger, I.~Malavolta, L.~Pascarella, F.~Palomba, D.~Di~Nucci, and
  A.~Bacchelli, ``A graph-based dataset of commit history of real-world android
  apps,'' in \emph{{Int. Conference on Mining Software Repositories (MSR)}},
  2018, pp. 30--33.

\bibitem{coppola2019migrationkotlin}
R.~Coppola, L.~Ardito, and M.~Torchiano, ``Characterizing the transition to
  kotlin of android apps: A study on f-droid, play store, and github,'' in
  \emph{{Int. Workshop on App Market Analytics (WAMA)}}, 2019, pp. 8--14.

\bibitem{Fdroid}
F-Droid, ``F-droid: software repository for android,''
  \url{https://www.f-droid.org/}.

\bibitem{UserGuide}
\BIBentryALTinterwordspacing
``Android user guide.'' [Online]. Available:
  \url{http://googlesamples.github.io/android-custom-lint-rules/user-guide.html}
\BIBentrySTDinterwordspacing

\bibitem{ApiGuide}
\BIBentryALTinterwordspacing
``Android lint api guide.'' [Online]. Available:
  \url{http://googlesamples.github.io/android-custom-lint-rules/api-guide.md.html}
\BIBentrySTDinterwordspacing

\bibitem{Rosner2011}
B.~Rosner, \emph{Fundamentals of Biostatistics}, 7th~ed.\hskip 1em plus 0.5em
  minus 0.4em\relax Brooks/Cole, Boston, MA, 2011.

\end{thebibliography}
